\documentclass{ims9x6}
\usepackage[USenglish]{babel} 
\usepackage{graphicx}
\usepackage{cite}
\usepackage{multirow}
\usepackage{placeins}
\usepackage{amsmath}
\usepackage{curves}
\usepackage[T1]{fontenc} 

\newcommand{\D}{\mathrm{d}}                       
\providecommand*{\I}{\mathrm{i}}                  
\newcommand{\tr}{\mathrm{tr}}			  
\newcommand{\ds}{\displaystyle}                   
\newcommand{\nn}{\nonumber}                       
\newcommand{\scaledmath}[2]{\scalebox{#1}{$\ds{#2}$}} 

\newcommand{\mw}[1]{\left\langle\right.\hspace{-0.5ex}{#1}%
                    \left.\hspace{-0.5ex}\right\rangle}%
\newcommand{\bok}[3]{\left\langle\right.\hspace{-0.5ex}{#1}%
                     \left.\hspace{-0.5ex}\right|{#2}\left|\right.%
                     \hspace{-0.5ex}{#3}\left.\hspace{-0.5ex}\right\rangle}
\newcommand{\e}[1]{\mathrm{e}^{#1}}     
\newcommand{\W}{_{\mathrm{W}}}          

\newcommand{\Int}[1][-5pt]{\int\limits_{\begin{picture}(16,3)(-8,-3)%
		\put(0,0){\curve(-3,0,-8,0)\curve(3,0,8,0)}%
		\put(8,0){\curve(0,0,-1.5,1.5)\curve(0,0,-1.5,-1.5)}%
		\put(0,0){\arc(-3,0){180}}\put(0,0){\makebox(0,0){$\cdot$}}%
		\end{picture}}\hspace*{#1}}

\DeclareMathAlphabet{\vecfont}{OT1}{cmr}{bx}{it}
\renewcommand{\vec}[1]{\vecfont{#1}}

\newcommand{\nab}{\boldsymbol{\nabla}}

\makeatletter
\renewcommand{\ps@plain}{%
  \renewcommand{\@oddhead}{\hfil\footnotesize%
    \raisebox{30pt}[0pt][0pt]{\parbox{300pt}{\centering%
      A contribution to the Proceedings of the\\{}%
      Workshop on Density Functionals for Many-Particle Systems\\{}
      2--29 September 2019, Singapore}}\hfil}%
  \renewcommand{\@evenhead}{\@oddhead}%
  \renewcommand{\@oddfoot}{\hfil\footnotesize%
        \raisebox{-8pt}[0pt][0pt]{\thepage}\hfil}%
  \renewcommand{\@evenfoot}{\@oddfoot}%
}
\makeatother
\renewcommand{\trimmarks}{\relax}

\begin{document}

\chapter{\uppercase{Density-potential functional theory\\%
                                   for fermions in one dimension}}

\markboth{M.-I. Trappe, J. H. Hue, and B.-G. Englert}%
{Density-potential functional theory for fermions in one dimension}

\author{Martin-Isbj\"orn~Trappe,$^{a}$ %
        Jun Hao Hue,$^{a,b}$%
        and Berthold-Georg~Englert$^{a,c,d}$}
        
\address{$^a$Centre for Quantum Technologies, %
             National University of Singapore, Singapore\\%
         $^b$Graduate School for Integrative Sciences \& Engineering, \\%
	     National University of Singapore, Singapore\\%
         $^c$Department of Physics, %
             National University of Singapore, Singapore\\%
         \makebox[0pt][c]{$^d$MajuLab, CNRS-UCA-SU-NUS-NTU International %
                          Joint Research Unit, Singapore}
         \\[1ex]%
         martin.trappe@quantumlah.org, %
         junhao.hue@u.nus.edu, %
         cqtebg@nus.edu.sg}

\begin{abstract}
We showcase the advantages of orbital-free density-potential functional theory
(DPFT), a more flexible variant of Hohenberg--Kohn density functional theory.
DPFT resolves the usual trouble with the gradient-expanded kinetic energy
functional by facilitating systematic semiclassical approximations in terms of
an effective potential energy that incorporates all interactions.
With the aid of two systematic approximation schemes we demonstrate that DPFT
is not only scalable, universally applicable in both position and momentum
space, and allows kinetic and interaction energy to be approximated
consistently, but can also compete with highly accurate, yet restricted,
methods.
As two- and three-dimensional geometries are extensively covered elsewhere,
our focus here is on one-dimensional settings, with semiclassical observables
systematically derived from both the Wigner function formalism and a
split-operator approach.
The high quality of our results for Fermi gases in Morse potentials invites
the use of DPFT for describing more exotic systems, such as trapped large-spin
fermion mixtures with contact or dipole-dipole interactions. 
\end{abstract}

\section{Introduction}\label{intro}
A gradient-free semiclassical particle density with enormous improvement over
the Thomas--Fermi (TF) approximation \cite{Thomas1927,Fermi1927} has recently
been developed for one-dimensional (1D) fermionic systems and applied to
particle numbers as small as two \cite{Ribeiro2015,Ribeiro2017,Ribeiro2018}
--- a regime that is usually regarded as out of reach for the TF model.
Ribeiro~\textit{et~al.} in Ref.~\cite{Ribeiro2015} deliver a systematic
correction to TF (which we subsequently refer to as `RB15'):
Despite the many approximations employed in arriving at the final expressions
for particle density and kinetic energy, the derivations are void of ad hoc
manipulations.
Naturally, a systematic and scalable methodology beyond the TF approximation
for systems in 1D, 2D, and 3D, which is both accurate and computationally
efficient, would enjoy wide-spread application.
However, RB15 is ultimately based on matching Langer-corrected WKB wave
functions \cite{Langer1937}, such that the extension to 2D and 3D does not
seem to be possible \cite{Burkeprivcom}.
That should not distract, however, from the profound insight that
``semiclassics is the hidden theory behind the success of DFT''
\cite{Okun2021}.  

Density-potential functional theory (DPFT) presents a unified framework for
handling corrections to the TF approximation.
The ground work for orbital-free DPFT was laid in the 1980s
\cite{Berge1982,Berge1984,BGE1984,BGE1984b,Berge1985,Berge1985b,%
  Berge1988alternative,Englert2019}
with the introduction of an effective potential alongside the particle
density.
The standard density-only energy-functional of Hohenberg--Kohn density
functional theory \cite{HohenbergKohn1964,Dreizler1990alternative} is thereby
transformed into a still exact, yet more flexible variant.
Most importantly, the usual nightmare with gradient expansions of the kinetic
energy in orbital-free DFT is avoided altogether (see the kinetic energy
densities in Table~\ref{IntroTable}, as functions of the one-particle density
$n(\vec r)$ at position~$\vec r$).
Moreover, approximations of kinetic energy and interaction energy can be made
consistent.
The latter trait is, for example, absent in Kohn--Sham (KS) DFT, where the
interaction energy is subjected to your favorite approximation while the
kinetic energy is exact.

\begin{table}[t]\centering
\tbl{The familiar leading gradient correction for dimension ${D=3}$, and its
  troublesome counterparts for ${D=1,2}$.
  We display the ground-state kinetic-energy densities
  $\tau^{\mathrm{TF}}[n]$ in Thomas--Fermi approximation and the leading
  corrections ${\tau^{\mathrm{sc}}[n]-\tau^{\mathrm{TF}}[n]}$ to order
  $\nab^2$ as presented in the literature.
  $m$ is the particle mass, $g$ is the spin multiplicity.%
  \rule{227pt}{0pt}}{
\begin{tabular}{c|ccc}\hline\hline\rule{0pt}{10pt}%
  $D$ & $\tau^{\mathrm{TF}}[n]$ & $\tau^{\mathrm{sc}}[n]-\tau^{\mathrm{TF}}[n]$
  & References\\[0.2em] \hline\rule[-9pt]{0pt}{25pt}%
  1 & \scaledmath{0.9}{\frac{\pi^2\hbar^2}{6m\,g^2}\bigl(n(x)\bigr)^3}
   & \scaledmath{0.9}{-\frac{\hbar^2}{24m}%
     \frac{\left(\partial_xn(x)^{\phantom{X}\hspace{-1.7ex}}\right)^2}
     {n(x)}}
  & \cite{Holas1991,Salasnich2007,Koivisto2007}\\[0.4em]
  \hline\rule{0pt}{15pt}%
  \multirow{2}{*}{2}
      & \multirow{2}{*}{\scaledmath{0.9}{%
        \frac{\pi\hbar^2}{m\,g}\bigl(n(\vec{r})\bigr)^2}}
  & \scaledmath{0.9}{\frac{\hbar^2}{24m}\,\delta\bigl(n(\vec{r})\bigr)
    \bigl(\nab n(\vec{r})\bigr)^2}
  & \cite{Brack2003alternative,DissvanZyl}\\[0.2em]
  & & 0
  & \hspace{-0.5ex}\cite{Holas1991,Salasnich2007,Koivisto2007,Putaja2012}\\
    \hline\rule[-9pt]{0pt}{25pt}%
  3 & \scaledmath{0.9}{\frac{\hbar^2}{5m}%
      \left(\frac{3^5\pi^4}{2g^2}\right)^{\hspace{-0.3ex}1/3}%
     \bigl(n(\vec{r})\bigr)^{5/3}}
   & \scaledmath{0.9}{\frac{\hbar^2}{72m}
     \frac{\left(\nab n(\vec{r})^{\phantom{X}\hspace{-1.7ex}}\right)^2}
     {n(\vec{r})}}
  & \hspace{-0.5ex}
    \cite{Holas1991,Salasnich2007,Koivisto2007,Kirzhnits1957}\\
  \hline\hline
\end{tabular}\label{IntroTable}}
\end{table}

The purpose of the present article is two-fold:
First, we introduce DPFT and extend our previous works on two- and
three-dimensional geometries \cite{Trappe2016,Trappe2017,Chau2018,TrHoAd2019}
to the 1D setting.
Specifically, we illustrate the workings of DPFT within two disjunct
approaches: (i) we deploy systematic gradient expansions in the effective
potential using Wigner's phase space formalism, see
\cite{Trappe2016,Trappe2017} and references therein; (ii) we utilize a
split-operator approach to retrieve a hierarchy of approximations without a
gradient expansion, see \cite{Chau2018} and references therein. Second, we
derive and benchmark densities and energies against the exact values for up to
1000 noninteracting fermions in a 1D Morse potential.
These examples demonstrate that DPFT can deliver accuracies of energies and
particle densities that are of similar quality as those obtained from RB15
\cite{Ribeiro2015,Ribeiro2017,Ribeiro2018}.
While RB15 works for one-dimensional problems only, there is nothing special
about 1D from the DPFT perspective, where particle densities and energies can
be derived within various frameworks in a unified way for all geometries.

\section{Density-potential functional theory in a nutshell}\label{DPFT}
We start from Hohenberg--Kohn DFT, which aims at finding the extremum of the
density functional of the total energy 
\begin{equation}\label{gsEnergy2}
E=E[n,\mu]=E_{\mathrm{kin}}[n]+E_{\mathrm{ext}}[n]+E_{\mathrm{int}}[n]
+\mu\left(N-\int(\D\vec r)\,n(\vec r)\right)
\end{equation}
(composed of kinetic, external, and interaction energy) of a system
constrained to $N$ particles via the chemical potential $\mu$.
We introduce the effective potential energy by
\begin{equation}\label{deltaEkin}
V(\vec r)=\mu-\frac{\delta E_{\mathrm{kin}}[n]}{\delta n(\vec r)}
\end{equation}
and write
\begin{equation}\label{LegendreTF}
  E_1[V-\mu]=E_{\mathrm{kin}}[n]
             +\int(\D\vec r)\,\bigl(V(\vec r)-\mu\bigr)\,n(\vec r)
\end{equation}
for the Legendre transform of the kinetic energy functional
$E_{\mathrm{kin}}[n]$, such that 
\begin{align}\label{EnergyVnmu}
E=&\;E[V,n,\mu]\nn\\
  =&\;E_1[V-\mu]-\int(\D\vec r)\,n(\vec r)\,\bigl(V(\vec r)
     -V_{\mathrm{ext}}(\vec r)\bigr) +E_{\mathrm{int}}[n]+\mu N\,.
\end{align}
At the stationary point of $E[V,n,\mu]$, the $V$- and $n$-variations obey
\begin{align}
  \delta V: & \qquad \,n[V,\mu](\vec r)=
              \frac{\delta E_1[V-\mu]}{\delta V(\vec r)}\label{nasdef}
\intertext{and}
 \delta n: & \qquad\label{Vdef}
          V[n](\vec r)=V_{\mathrm{ext}}(\vec r)
          +\frac{\delta E_{\mathrm{int}}[n]}{\delta n(\vec r)}\,,
\end{align}
respectively.
The $\mu$-variation, combined with Eq.~(\ref{nasdef}), reproduces the
par\-ticle-number constraint  
\begin{equation}\label{PartNumConstraint}
\int(\D\vec r)\,n(\vec r)=N\,.
\end{equation}
Equation~(\ref{nasdef}) immediately yields the particle density in the
noninteracting case (${V=V_{\mathrm{ext}}}$).
For a given value of $\mu$, $n$ is a functional of $V$, and vice versa, such
that a self-consistent solution of
Eqs.~(\ref{nasdef})---(\ref{PartNumConstraint}) in the case of interacting
systems produces the ground-state density, much like in the KS scheme, but
without resorting to orbitals.
Details of the numerical implementation of the self-consistent loop for DPFT
can be found in Appendix~E of Ref.~\cite{TrHoAd2019}.
Further details and implications of
Eqs.~(\ref{EnergyVnmu})--(\ref{PartNumConstraint}) are well documented in the
literature, see
\cite{Berge1988alternative,Trappe2016,Trappe2017,Berge1992}, for example. 

While $E_{\mathrm{kin}}[n]$ is not known explicitly, it is a simple exercise
to find the single-particle trace 
\begin{equation}
E_1[V-\mu]=\tr\bigl(\mathcal E_T(H-\mu)\bigr)\,,\label{tracef}
\end{equation}
for independent particles, see Refs.~\cite{Trappe2017,Berge1992}, with the
ground-state version recovered as the temperature $T$ tends to zero.
Here, the operator\footnote{We shall omit arguments of functions for brevity
  wherever the command of clarity permits.} 
\begin{equation}\label{fHmuT}
  \mathcal E_T(A=H-\mu)
  =(-k_{\mathrm{B}}T)\,\ln\Bigl(1+\mathrm{e}^{-A/(k_{\mathrm{B}}T)}\Bigr)
\end{equation}
is a function of the single-particle Hamiltonian
\begin{equation}\label{singlePartHamil}
H(\vec R,\vec P)=\frac{\vec P^2}{2m}+V(\vec R)\,,
\end{equation}
where $\vec R$ and $\vec P$ are the single-particle position and momentum
operators, respectively.
Equation~(\ref{tracef}) and its zero-temperature limit\footnote{For ${T\to0}$
  we obtain ${\mathcal E_0(H-\mu)=(H-\mu)\,\eta(\mu-H)}$, where $\eta(\ )$ is
  the step function.}
are exact for noninteracting systems and can be made exact for interacting
systems if the interacting part of the kinetic-energy density-functional is
transferred from Eq.~(\ref{LegendreTF}) to $E_{\mathrm{int}}[n]$. 

Any approximation of the single-particle trace in Eq.~(\ref{tracef}) yields a
corresponding approximation for the particle density in Eq.~(\ref{nasdef}). 
We are now in the position to benchmark semiclassical approximations of $E_1$
unambiguously for systems with a known interaction functional and for
noninteracting systems.  

We shall employ and explore two independent approximation schemes for DPFT.
First, as worked out in Ref.~\cite{Trappe2017}, we express the trace in
Eq.~(\ref{tracef}) as the classical phase space integral\footnote{The trace
  includes the degeneracy factor $g$.} 
\begin{equation}\label{trH}
  \tr\Bigl(\mathcal E_T\bigl(A(\vec R,\vec P)\bigr)\Bigr)
  =g\int\frac{(\D\vec r)(\D\vec p)}{(2\pi\hbar)^D}\,
  \bigl[\mathcal E_T(A)\bigr]\W(\vec r,\vec p)\,,
\end{equation}
where the momentum integral over the Wigner function $\big[\mathcal
E_T(A)\big]\W$ of $\mathcal E_T(A)$ can be approximated
by\footnote{${\mw{f}_\mathrm{Ai}=%
    \mathop{\scaledmath{0.6}{\int}}\limits_{-\infty}^\infty\!%
    \D x\,\mathrm{Ai}(x)f(x)}$, %
  with the Airy function $\mathrm{Ai}(\ )$, %
  denotes the Airy average of the function $f(x)$, %
  and `$\cong$' stands for an approximation that reproduces %
  the leading correction exactly.} 
\begin{equation}
  \int(\D\vec p)\,\big[\mathcal E_T( A)\big]\W(\vec r,\vec p)
  \cong\int(\D\vec p)\,{\left\langle\!\mathcal E_T\bigl(\tilde{A}\W\bigr)
  -\frac{\hbar^2\left(\nab^2V\right)}{12m}\mathcal E_T''
  \bigl(\tilde{A}\W\bigr)\!\right\rangle}_{\hspace*{-0.2em}\mathrm{Ai}}.
  \rule[-18pt]{0pt}{5pt}\label{trfA4}
\end{equation}
Here,
\begin{equation}\label{AWtilde}
\tilde{A}\W(\vec r,\vec p)=H\W(\vec r,\vec p)-\mu-x\, a(\vec r)\,,
\end{equation}
with the Wigner function ${H\W(\vec r,\vec p)=\vec p^2/(2m)+V(\vec r)}$ of the
single-particle Hamiltionian 
and ${a(\vec r)=|\hbar\nab V(\vec r)|^{2/3}/(2m^{1/3})}$.
Equation (\ref{trfA4}) is exact up to the leading gradient correction
$\left[\mathcal O\left(\nab^2\right)\right]$, and thus presents a systematic
correction to the TF approximation, which is recovered in the uniform limit
$\left[\mathcal O\left(\nab^0\right)\right]$.
However, the `Airy-average' in Eq.~(\ref{trfA4}) contains also higher-order
gradient corrections that enter through the Moyal products from the power
series expansion of $\mathcal E_T(A)$ in Eq.~(\ref{trH}).
These higher-order corrections are responsible for the almost exact behavior
of particle densities across the boundary between classically allowed and
forbidden regions, where the TF approximation can fail epically (even if
supplemented with the leading gradient correction \cite{Trappe2017}).
We obtain the approximate particle densities for one-, two-, and
three-dimensional geometries by combining Eqs.~(\ref{nasdef}) and
(\ref{tracef})--(\ref{AWtilde}) and evaluating the momentum integral of
Eq.~(\ref{trfA4}).
The 2D case is covered extensively in Refs.~\cite{Trappe2016,Trappe2017}, and
a manuscript that covers the 3D formulae, in particular for computational
chemistry, is in preparation \cite{TrappeWittManzhos2020}.
Here, we focus on the 1D situation and provide details in Secs.~\ref{sec1D}
and \ref{secEnergies}. 

Our second approximation scheme reiterates some of the results in
Ref.~\cite{Chau2018} and begins with realizing that Eqs.~(\ref{nasdef}) and
(\ref{tracef}) at ${T=0}$ yield\footnote{In Eq.~(\ref{nSTA}), we make use of
  the Fourier transform of the step function $\eta(\ )$, and the integration
  path from ${t=-\infty}$ to ${t=\infty}$ crosses the imaginary $t$ axis in
  the lower half-plane.} 
\begin{equation}
n(\vec r)=g\bok{\vec r}{\eta(\mu-H)}{\vec r}
         =g\Int\frac{\D t}{2\pi\I t}\,\e{\frac{\I t}{\hbar}\mu}\,
           \bok{\vec r}{U(t)}{\vec r}\,,\label{nSTA}
\end{equation}
which invites tailored Suzuki--Trotter (ST) factorizations of the unitary
time-evolution operator ${U(t)=\e{-\frac{\I t}{\hbar}H}}$.
We obtain a hierarchy of approximations of $n(\vec r)$ from appropriate
coefficients $\alpha_i$ and $\beta_i$ in the ansatz%
\footnote{The choice ${\nu=1,\;\alpha_1=\beta_1=1}$ neglects the
  noncommutativity of $\vec R$ and $\vec P$ and delivers the TF
  density; $U_2(t)$ recovers $U(t)$ up to $\mathcal O(t)$. $\lceil\ \rceil$ is
  the integer-valued ceiling function.}  
\begin{equation}\label{Unu}
  U(t)\approx U_\nu(t)
  =\prod_{i=1}^{\lceil \nu/2 \rceil} \e{- \frac{\I t}{\hbar} \alpha_i
    V(\vec{R})}\, \e{- \frac{\I t}{\hbar} \beta_i \vec{P}^2/(2m)}\,,
\end{equation}
where the exponential factors are multiplied from left to right in order of
increasing $i$ values. The choice ${\alpha_1=0}$, ${\alpha_2=1}$, and
${\beta_1=\beta_2=1/2}$ yields 
\begin{equation}\label{n3p}
  n_{3'}(\vec r)=g\int(\D\vec a)\left(\frac{k_{3'}}{2\pi a}\right)^D
  \mathrm{J}_D\bigl(2ak_{3'}\bigr)\,,
\end{equation}
with
$k_{3'}=\frac{1}{\hbar}\big[2m\,\bigl(\mu-V(\vec r+\vec a)\bigr)\big]_+^{1/2}$,
${[z]_+=z\,\eta(z)}$, and $\mathrm{J}_D(\ )$ the Bessel function of order $D$.
Equation~(\ref{n3p}) is a first nonlocal correction to the TF density, since
$U_{3'}(t)$ recovers $U(t)$ up to $\mathcal O(t^2)$. 
Reference \cite{Chau2018} also features an approximation $n_7$ for the
particle density\footnote{$n_7$ is denoted as
  $\left.n_7\right|_{\epsilon\to0}$ in Ref.~\cite{Chau2018}.
  The underlying approximation $\left. U_7\right|_{\epsilon\to 0}$ consists of
  seven $\epsilon$-dependent exponential factors that reduce to the five
  factors shown in Eq.~(\ref{Unu'}) for ${\epsilon\to 0}$ and has been
  proven highly accurate for a variety of systems
  \cite{Chin1997,Omelyan2002,Chin2005,Chau2018,Hue2020,Paraniak2021}.}
based on merely five exponential factors, 
\begin{align}\label{Unu'}
 & \alpha_1=\alpha_3=\frac{1}{6}\,,\quad
  \alpha_2 V=\frac{2}{3}V-\frac{1}{72m}[t\nab V]^2\,,\nn\\
& \beta_1=\beta_3=\frac{1}{2}\,,\quad\beta_2=0\,,
\end{align}
instead of the eleven factors needed for reaching the same order of accuracy
$\left[\mathcal O\left(t^4\right)\right]$ if $\alpha_i$ and 
$\beta_i$ are required to be constants \cite{Hatano2005}:
Equation~(\ref{nSTA}) with Eqs.~(\ref{Unu}) and (\ref{Unu'}) yields
\begin{equation}\label{n7}
n(\vec r)\cong n_7(\vec{r})
={\left\langle g\int(\D\vec{a})\,
    {\left(\frac{k^{(x)}_{7}}{2\pi a}\right)}^D
    \mathrm{J}_D{\left(2ak^{(x)}_{7}\right)}
  \right\rangle}_{\hspace{-0.2em}\mathrm{Ai}}\,,
\end{equation}
where
\begin{align}
  k^{(x)}_{7}&=\frac{1}{\hbar}
               \Biggl[2m{\left(\mu-\frac{1}{3}\bigl[V(\vec{r})
               +2V(\vec{r}+\vec{a})\bigr]\right)}\nn\\
  &\hphantom{=\frac{1}{\hbar}\Biggl[}\mbox{}-x
    {\left(\frac{1}{3}\bigl[\hbar m\nab V(\vec{r}+\vec{a})\bigr]^2
               \right)}^{1/3}\Biggr]^{1/2}_+\,.
\end{align}
Since $n_7$ is exact up to fourth order
in $t$, it is exact for the linear potential and includes the leading gradient
correction.
In contrast to the local TF density, $n_{3'}(\vec r)$ and $n_7(\vec r)$ are
fully nonlocal expressions, which sample the effective potential $V$ in a
neighborhood of the position $\vec r$, and can provide highly accurate density
profiles, as tested for 3D in Ref.~\cite{Chau2018}.
We assess the quality of Eqs.~(\ref{n3p}) and (\ref{n7}) for ${D=1}$ in
Sec.~\ref{results} below.

\section{Airy-averaged densities in 1D}\label{sec1D}
The semiclassical particle densities from the Wigner function approach in 1D
and 3D can be obtained analogously to the 2D case covered in
Ref.~\cite{Trappe2017}.
We now focus on the 1D situation, for which the so far available orbital-free
gradient corrections of the kinetic energy are not bounded from below, see
Table~\ref{IntroTable}.
The semiclassical closed expression RB15 for the particle density
\cite{Ribeiro2015} proves to be remarkably accurate even for two particles,
and also in the vicinity of the classical turning points, where the failure of
the TF approximation is most pronounced.
We will show in the following that, like in 2D, the Airy-averaged densities of
our DPFT formalism closely resemble the exact densities everywhere for very
low but finite temperatures.
Even for ${N=2}$ our Airy-averaging method delivers a high accuracy. 

As outlined in Sec.~\ref{DPFT}, Eq.~(\ref{nasdef}) eventually yields
\begin{equation}\label{nAi1D}
  n(r,T)\cong n^{\mathrm{Ai}}(r,T)
  =g\frac{\sqrt{2m}}{\pi\hbar}\mw{b_x(r)\, l_x(r,T)}_{\mathrm{Ai}}
\end{equation}
for ${V'\not=0}$, where
\begin{align}
b_x(r)&=1-\frac{\hbar^2 U''(r)}{12m\,a(r)^2}x \,,\nn\\
  l_x(r,T)&=-\frac{\sqrt{\pi k_{\mathrm{B}}T}}{2}\mathrm{Li}_{1/2}
    \Bigl(-\mathrm{e}^{\left[x\,a(r)-U(r)\right]/(k_{\mathrm{B}}T)}\Bigr),\nn\\
l_x(r,0)&=\sqrt{a(r)}\,\sqrt{x-y(r)}\,\eta\bigl(x-y(r)\bigr)\,,\nn\\
y(r)&=U(r)/a(r)\,,\nn\\
U(r)&=V(r)-\mu\,,
\end{align}
${U'(r)=\partial_rU(r)}$, ${U''(r)=\partial_r^2U(r)}$, and
$\mathrm{Li}_{1/2}(\ )$ is the polylogarithm of order $1/2$.
With ${\tilde{A}\W\to A\W}$ for ${V'\to 0}$, the Airy-average in
Eq.~(\ref{trfA4}) drops out, and we find\footnote{The numerical evaluation of
  the momentum integral in Eq.~(\ref{nAiTVp0}) poses no difficulties.} 
\begin{align}\label{nAiTVp0}
  n^{\mathrm{Ai}}\bigl(\left.r\right|_{V'=0},T\bigr)
  &=-\frac{g}{\hbar}\sqrt{\frac{m\,k_{\mathrm{B}}T}{2\pi}}
    \mathrm{Li}_{1/2}\Bigl(-\mathrm{e}^{-U/(k_{\mathrm{B}}T)}\Bigr)\nn\\
  &\phantom{=}-\frac{g\,\hbar\,U''}{96\,\pi\,m\,(k_{\mathrm{B}}T)^2}
    \int\D p\,\frac{\sinh\bigl(\gamma(p)\bigr)}
    {\bigl[\cosh\bigl(\gamma(p)\bigr)\bigr]^3}\,,
\end{align}
where ${\gamma(p)=A\W/(2k_{\mathrm{B}}T)}$.

The zero-temperature limit of the density expression in Eq.~(\ref{nAi1D})
reads\footnote{$\mathrm{sgn}(\ )$ denotes the signum function and
  $\mathrm{Ai}'(\ )$ the derivative of the Airy function.} 
\begin{align}\label{CompareBurkenAi}
n^{\mathrm{Ai}}(r,0)
  &=g\frac{p_\mathrm{F}}{\hbar}\Biggl[-\mathrm{sgn}(\bar{y})\,
    \sqrt{|\bar{y}|}\,\mathrm{Ai}(\bar{y})^2
    +\frac{\mathrm{Ai}'(\bar{y})^2}{\sqrt{|\bar{y}|}}\nn\\
  &\hphantom{=g\frac{p_\mathrm{F}}{\hbar}\Biggl[}
    +\frac{\hbar\,U''}{3\,p_\mathrm{F}\,|U'|}
    \mathrm{Ai}(\bar{y})\mathrm{Ai}'(\bar{y})\Biggr]\,,
\end{align}
with the local Fermi momentum ${p_{\mathrm{F}}=\sqrt{2m|U(r)|}}$ and
${\bar{y}=2^{-2/3}y(r)}$.
The Airy-averaged ground-state density in Eq.~(\ref{CompareBurkenAi}) holds
for ${U\not=0}$ (and ${U'\not=0}$) and exhibits a striking structural
resemblance to the expression of the semiclassical particle density RB15 given
in Ref.~\cite{Ribeiro2015}, although the latter is void of gradients of the
potential.
While beyond the scope of this article, we deem it a worthy enterprise to
explore the deeper connections between RB15 and Eq.~(\ref{CompareBurkenAi}).
This may possibly lead to a cure for the unphysical oscillations of
$n^{\mathrm{Ai}}(r,0)$ near the extremal points of $U$, see
Fig.~\ref{Figures_DFMPS2019_TrappeHueEnglert_Morse_N2_nAi} below.

\section{DPFT densities for the Morse potential}\label{results}
For our benchmarking exercise we refrain from considering the harmonic
oscillator potential, for which the TF energy of noninteracting spin-polarized
(${g=1}$) fermions in 1D accidentally coincides with the exact energy (for
integer $N$).
As in Ref.~\cite{Ribeiro2015}, we rather opt for the Morse
potential\footnote{We declare $\hbar\omega$ and $\sqrt{\hbar/(m\omega)}$
  (`[osc]') the units for energy and length, respectively, and set
  ${\hbar=m=\omega=1}$.} 
\begin{align}\label{Morse}
V_\mathrm{ext}(r)=V_0\,\left(e^{-r/2} - 2 e^{-r/4}\right)
\end{align}
of depth $V_0$.

\begin{figure}[t]
\begin{center}
\includegraphics[width=0.81\linewidth]{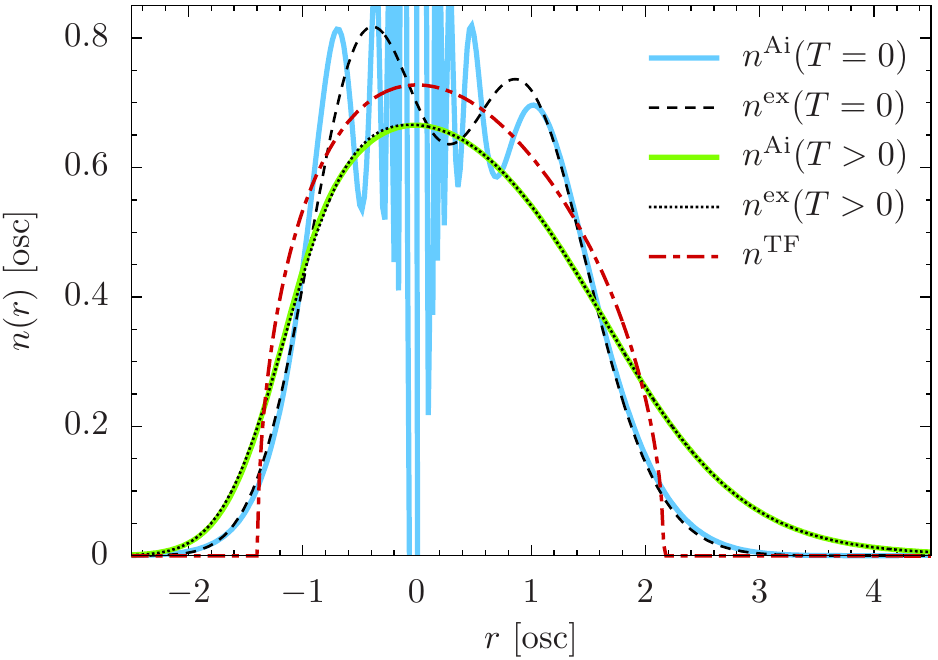}
\caption{\label{Figures_DFMPS2019_TrappeHueEnglert_Morse_N2_nAi}%
  Particle densities for the 1D Morse potential
  in Eq.~(\ref{Morse}) with ${V_0=15}$.
  While $n^{\mathrm{Ai}}(r,T>0)$ (${k_{\mathrm{B}}T=1}$) agrees well with the
  exact density $n^{\mathrm{ex}}(r,T>0)$ everywhere, in particular near the
  classical turning points, $n^{\mathrm{Ai}}(r,0)$ exhibits unphysical
  oscillations that are due to the second line of Eq.~(\ref{CompareBurkenAi}),
  which diverges as ${U'\to 0}$.
  For the 2D case, similar oscillations are linked to an ill-defined
  zero-temperature limit, see \protect\cite{Trappe2017}. 
  RB15 from Ref.~\protect\cite{Ribeiro2015} essentially coincides with
  ${n^{\mathrm{ex}}(r,T=0)}$.}
\end{center}
\end{figure}

\begin{figure}[t]
\begin{center}
\includegraphics[width=0.81\linewidth]{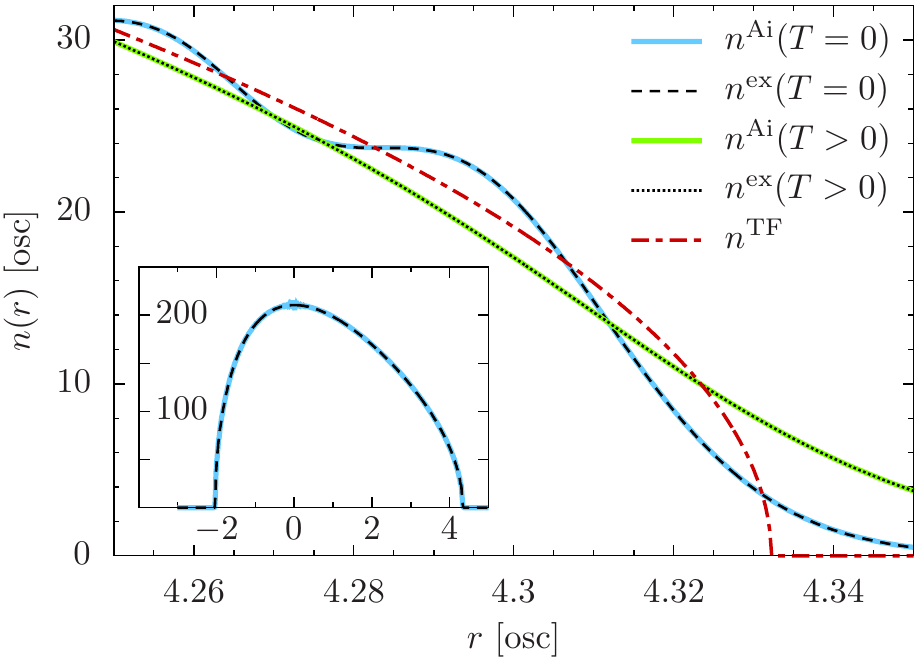}
\caption{\label{Figures_DFMPS2019_TrappeHueEnglert_Morse_N1000_nAi}%
  Particle densities as in
  Fig.~\ref{Figures_DFMPS2019_TrappeHueEnglert_Morse_N2_nAi}, but for
  ${N=1000}$   fermions, with ${V_0=5\times10^5}$ and
  ${k_{\mathrm{B}}T=1000}$. 
  Evidently, both $n^{\mathrm{Ai}}(r,T)$ and $n^{\mathrm{Ai}}(r,0)$ leave
  nothing to be desired in the vicinity of the classical turning points (main
  plot).
  The inset shows that the unphysical oscillations of $n^{\mathrm{Ai}}(r,0)$
  near the minimum of the Morse potential are much better behaved for
  ${N=1000}$ than for ${N=2}$, which stresses the fact that $n^{\mathrm{Ai}}$
  is a semiclassical approximation.} 
\end{center}
\end{figure}

\begin{figure}
\begin{center}
\includegraphics[width=0.81\linewidth]{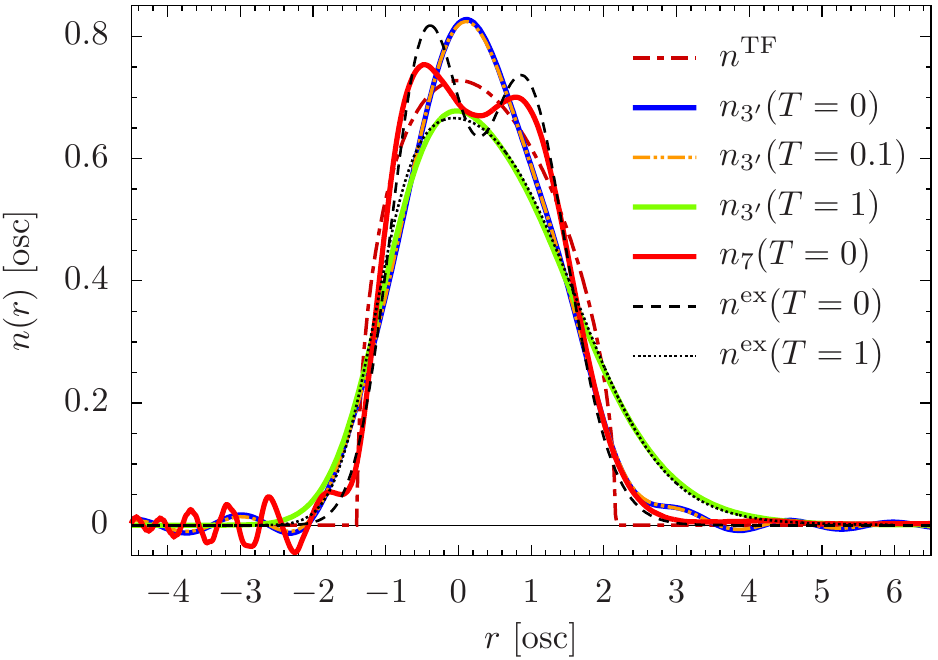}
\caption{\label{Figures_DFMPS2019_TrappeHueEnglert_Morse_N2_nnu}%
Compared with the Airy-averaged density $n^{\mathrm{Ai}}$ in
Fig.~\ref{Figures_DFMPS2019_TrappeHueEnglert_Morse_N2_nAi}, the density tails
and the finite-temperature profiles of $n_{3'}$ are less accurate, but they
are well-behaved at any temperature.
$n_7$ qualitatively captures the exact oscillations in the classically allowed
region but exhibits unphysical oscillations in the classically forbidden
region (at negative $r$), where large gradients of $V$ pose a challenge.}  
\end{center}
\end{figure}

\begin{figure}
\begin{center}
\includegraphics[width=0.81\linewidth]{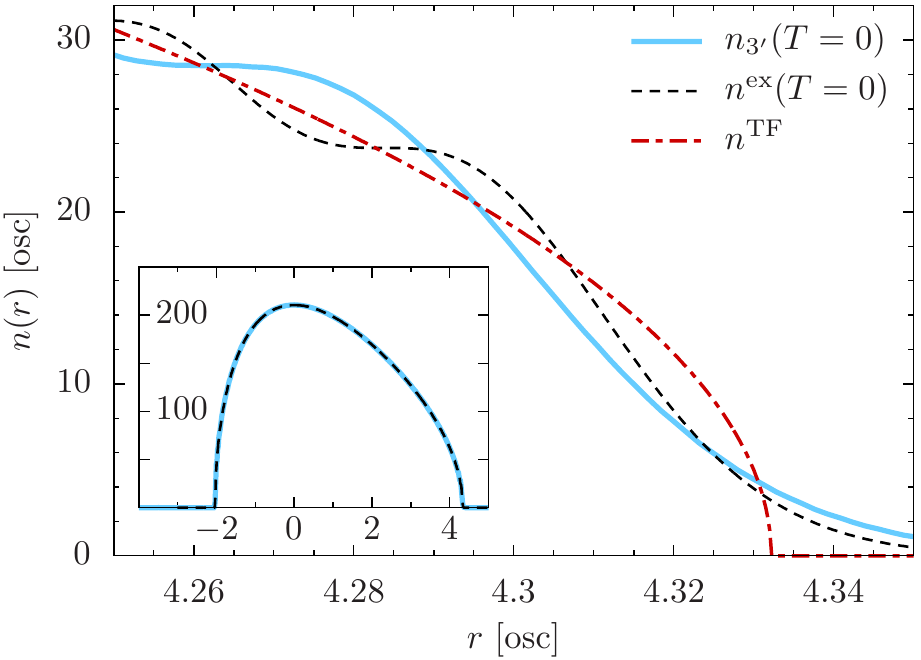}
\caption{\label{Figures_DFMPS2019_TrappeHueEnglert_Morse_N1000_n3p}%
Like Fig.~\ref{Figures_DFMPS2019_TrappeHueEnglert_Morse_N1000_nAi}, but with
the density formula $n_{3'}$ of Eq.~(\ref{n3p}).
As expected, $n_{3'}$ captures the exact density profile reasonably well
(inset), but does not match the oscillations into the classically forbidden
region (main plot) --- $n_{3'}$ is not exact for a linear potential, in
contrast to $n^{\mathrm{Ai}}$.}  
\end{center}
\end{figure}

\begin{figure}
\begin{center}
\includegraphics[width=0.81\linewidth]{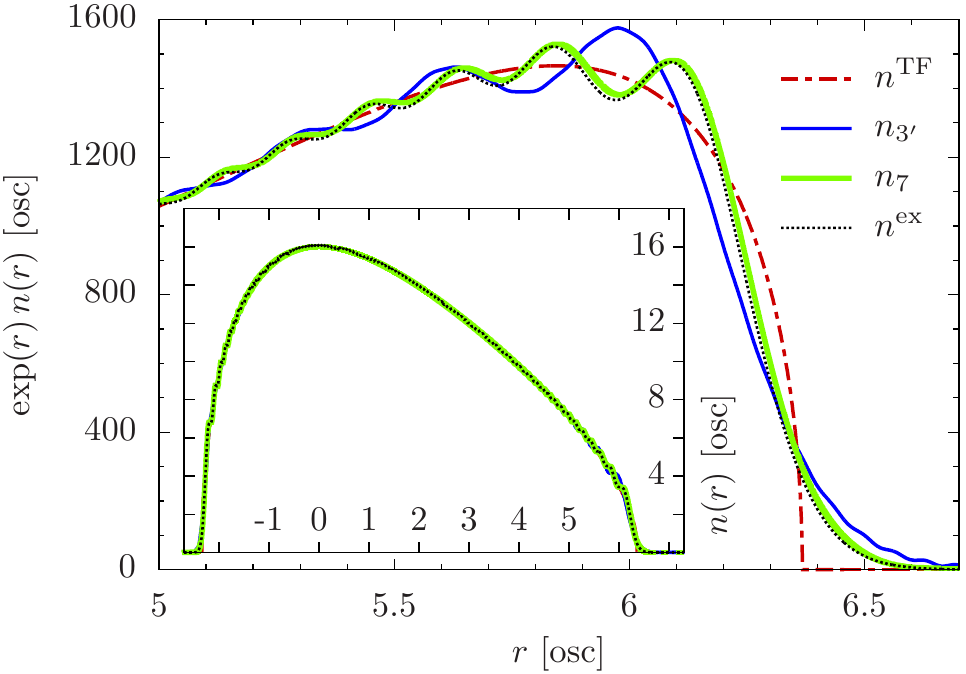}
\caption{\label{Figures_DFMPS2019_TrappeHueEnglert_Morse_N100_n7}%
In the semiclassical regime of ${N=100}$ fermions and ${V_0=2000}$, $n_7$
captures the oscillatory behavior of the exact density with excellent
accuracy.
The deviations from $n^{\mathrm{ex}}$ in the tail are only visible when
exponentially amplified (main plot). As expected, $n_7$ proves superior to
$n_{3'}$, whose oscillations are out of phase with the exact oscillations.}  
\end{center}
\end{figure}

Figure~\ref{Figures_DFMPS2019_TrappeHueEnglert_Morse_N2_nAi} illustrates the
high quality of our Airy-averaged 
densities for two fermions in the vicinity of the classical turning points,
where they are designed to perform well.
We compare the Airy-averaged expressions $n^{\mathrm{Ai}}(r,T)$ and
$n^{\mathrm{Ai}}(r,0)$ with the exact densities $n^{\mathrm{ex}}$
\cite{Frank2000} and the (zero-temperature) TF density $n^{\mathrm{TF}}$.
Bearing in mind that the TF approximation is not well suited for small
particle numbers, the corrections to TF included in $n^{\mathrm{Ai}}(r,T)$
deliver a remarkable accuracy.
As expected, this observation is even more striking for larger particle
numbers:
We demonstrate in
Fig.~\ref{Figures_DFMPS2019_TrappeHueEnglert_Morse_N1000_nAi} that our
Airy-averaged 
densities are indistinguishable (to the eye) from the exact densities near the
classical turning points for ${N=1000}$ fermions. 

\enlargethispage{0.5\baselineskip}

In the following, we assess the semiclassical density expressions
$n_{3'}(\vec r)$ and $n_7(\vec r)$, based on the Suzuki--Trotter
approximations in Eq.~(\ref{Unu}).
Figure~\ref{Figures_DFMPS2019_TrappeHueEnglert_Morse_N2_nnu} shows the density
of two noninteracting particles in the Morse potential with $n_{3'}$ of
Eq.~(\ref{n3p}) --- the simplest variant of $n_{\nu}$ beyond the TF
approximation --- and $n_7$ from Eq.~(\ref{n7}).
The derivation of the finite-temperature version ${n_{3'}(T>0)}$ is given
elsewhere \cite{Grochowski2021,TrappeWittManzhos2020}.
Oscillations around zero in the classically forbidden region appear for both
$n_{3'}$ and $n_7$.
That is, the higher-order errors of $n_\nu$ cannot be regarded small for some
effective potentials $V$ --- like the Morse potential whose gradient diverges
exponentially. However, these errors diminish for larger particle numbers:
Increasing the particle number to ${N=1000}$ in the deep Morse potential
${\left(V_0=5\times10^5\right)}$, we approach the quasiclassical regime, where
the TF density and its semiclassical corrections are largely indistinguishable
(to the eye) from the exact density, see
Fig.~\ref{Figures_DFMPS2019_TrappeHueEnglert_Morse_N1000_n3p}.
In Fig.~\ref{Figures_DFMPS2019_TrappeHueEnglert_Morse_N100_n7} we showcase the
quasi-exact behavior of $n_7$ for a moderate particle number of ${N=100}$, for
which the exact quantum oscillations are still visible to the eye and where
the semiclassical DPFT densities deliver, unsurprisingly, a much higher
accuracy compared with the case of ${N=2}$.

\section{Airy-averaged energies}\label{secEnergies}
We conclude with the benchmarking of the semiclassical energies of our Wigner
function approach.
The zero-temperature Airy-average of the potential functional $E_1[U]$ for
polarized Fermi gases in 1D is obtained from
Eqs.~(\ref{tracef})--(\ref{AWtilde}) and reads 
\begin{equation}\label{E1Airy1D}
  E_1^{\mathrm{Ai}}[U]
  =\int\D r\,{\left[\frac{U''}{6}{\left(\frac{\hbar^2}{2m|U'|}\right)}^{1/3}
    F_0(\bar{y})-\bigl|U'\bigr|F_2(\bar{y})\right]}\,,
\end{equation}
where ${F_2(\bar{y})=\frac23\bigl[\bar{y}^2\mathrm{Ai}(\bar{y})^2
  -\frac12\mathrm{Ai}(\bar{y})\mathrm{Ai}'(\bar{y})
  -\bar{y}\mathrm{Ai}'(\bar{y})^2\bigr]}$
and ${F_0(\bar{y})=\mathrm{Ai}(\bar{y})^2}$,
see Ref.~\cite{Berge1988alternative}.
While the oscillations of $n^{\mathrm{Ai}}(r,0)$ become singular as ${U'\to
  0}$, the integrand of $E_1^{\mathrm{Ai}}[U]$ remains finite throughout.
An analogous observation holds for the 2D case \cite{Trappe2017}.

We assess the performance of $E_1^{\mathrm{Ai}}[U]$ for the 1D Morse potential
in Eq.~(\ref{Morse}).
In Table~\ref{TableEnergies1D} we report the total energies for several
particle numbers~$N$. 
The Airy-averaged energies show a significant improvement over the TF energies
if the particle densities differ considerably from those of the harmonic
oscillator, that is, if the chemical potential is close to the bound-state
threshold.\footnote{The TF energies equal the exact energies in the case of a
  1D harmonic oscillator potential.
  It is therefore difficult to improve upon the TF energies if the Morse
  potential supports many more bound states than the chosen particle number
  $N$.
  Then the Morse potential is similar to the harmonic-oscillator potential in
  the spatial range relevant for the particle densities.}

\begin{table}[t]\centering
\tbl{The comparison of
  ${E^{\mathrm{Ai}}=E_1^{\mathrm{Ai}}+\mu^{\mathrm{Ai}}N}$, see
  Eqs.~(\ref{EnergyVnmu}) and (\ref{E1Airy1D}), with the exact energies
  $E^{\mathrm{ex}}$ (see Ref.~\protect\cite{Frank2000}) for the 1D Morse
  potential in Eq.~(\ref{Morse}) reveals a satisfactory performance of the
  Airy-averaged energy functional in Eq.~(\ref{E1Airy1D}).
  The improvement of $E^{\mathrm{Ai}}$ over the TF energy $E^{\mathrm{TF}}$ is
  most visible if $N$ is close to the number $N_{\mathrm{max}}$ of bound
  states supported by $V_0$, i.e., if the part of $V$ that is most relevant
  for the density profile is least harmonic.}{\tabcolsep=5pt%
\begin{tabular}{@{\thinspace}cccccccc@{\thinspace}}\hline\hline\rule{0pt}{9pt}%
$V_0$ & 0.25 & 2.5 & 15 & 15 & 15 & ${3.2\times10^4}$ & ${5\times10^5}$ \\
  $N$[$N_{\mathrm{max}}$]& 2[3]& 8[9] & 2[22]& 8[22] & 20[22] & 1000[1012]
                       & 1000[4000]\\
\hline\rule{0pt}{10pt}%
  $E^{\mathrm{ex}}$ & -0.2246 & -27.340 & -81.495 & -7.424
                  & -109.420 & -10793887.5 & -385416664 \\
  $E^{\mathrm{TF}}$ & -0.2298 & -27.345 & -81.516 & -7.445
                  & -109.472 & -10793890.1 & -385416667 \\
  $E^{\mathrm{Ai}}$ & -0.2231 & -27.288 & -81.459 & -7.423
                  & -109.416 & -10793887.4 & -385416656 \\ \hline\hline
\end{tabular}\label{TableEnergies1D}}
\end{table}

The value $E_{\mathrm{kin}}^{\mathrm{Ai}}$ of the Airy-averaged kinetic energy
at the stationary point of $E[V,n,\mu]$ can also be evaluated unambiguously.
Employing Eq.~(B.15) from Ref.~\cite{Trappe2017},
\begin{align}\label{specialeAWp2}
&\int(\D\vec p)\,\frac{\vec p^2}{2m}\,\big[f(A)\big]\W(\vec r,\vec p)\nn\\
  \cong&\int(\D\vec p)\,\frac{\vec p^2}{2m}
         {\left\langle f\bigl(\tilde{A}\W\bigr)
         -\frac{\hbar^2\left(\nab^2V\right)}{12m}\frac{D-1}{D}
         f''\bigl(\tilde{A}\W\bigr)\right\rangle}_{\mathrm{Ai}}\,,
\end{align}
which holds for any function $f$ that has a Fourier transform, we find
\begin{align}
  E_{\mathrm{kin}}^{\mathrm{Ai}}
  &=\mathrm{tr}{\left\{\frac{P^2}{2m}\eta(\mu-H)\right\}}\nn\\
  &=\frac{g}{2\pi\hbar}\int\D r\,\D p\,\frac{p^2}{2m}
    {\left\langle\eta\bigl(x\,a-U-p^2/(2m)\bigr)\right\rangle}_{\mathrm{Ai}}
   \nn\\ 
&=g\int\D r\,\frac{|U'|}{2}\,F_2(\bar{y})\label{EkinAi}
\end{align}
for ${D=1}$ after performing the momentum integral and the Airy average in
Eq.~(\ref{EkinAi}).
At the ground state itself, the energies in Table~\ref{TableEnergies1D} are
indeed reproduced by
${E_{\mathrm{kin}}^{\mathrm{Ai}}%
  +\int\D r\,V_{\mathrm{ext}}(r)\,n^{\mathrm{Ai}}(r,0)}$,
with ${V=V_{\mathrm{ext}}}$ in the noninteracting case.
In contrast to $E_1^{\mathrm{Ai}}[U]$, however, Eq.~(\ref{EkinAi}) holds at
the ground state only and can therefore not be used as a functional for
determining the stationary points of the total energy. 

\enlargethispage{-2.0\baselineskip}

\section{Concluding remarks}
This presentation puts a spotlight on foundations and recent advances of
density-potential functional theory.
We complemented our technical arsenal for two- and three-dimensional settings,
see Refs.~\cite{Trappe2017,Chau2018,TrHoAd2019} and references therein, by
deriving systematic corrections to the Thomas--Fermi approximation for both
densities and energies in one dimension.
Benchmarking against exact values for noninteracting fermions in a Morse
potential at both zero and finite temperature, we found that our
`Airy-averaged' Wigner function approach delivers highly accurate density
tails across the boundary of the classically allowed region and can improve
significantly upon the already very accurate Thomas--Fermi energies.
We also demonstrated that our most accurate split-operator-based density
formula ($n_7$) delivers the quantum oscillations of the exact densities.
Our results underscore the high quality of our semiclassical techniques when
used in conjunction with an accurate interaction functional.
In other words, the kinetic energy is handled accurately without orbitals.

Both approximation schemes presented here are universally applicable to one-,
two-, and three-dimensional systems.
They produce systematically quantum-corrected densities in both position and
momentum space at zero as well as finite temperature.
Our concrete implementations of density-potential functional theory thus may
save the day whenever other orbital-free or Kohn--Sham approaches cannot
deliver \emph{systematic} quantum corrections for realistic large-scale
systems.
In view of this assembly of properties, we anticipate that density-potential
functional theory will prove particularly strong relative to other techniques
in extracting ground-state properties of interacting multi-component Fermi
gases --- for instance, of the contact- and dipole-dipole-interacting types
\cite{Fang2011,Lu2012,Baranov2012,Aikawa2014b,Trappe2016a,Grochowski2017,%
  Baier2018},
including their one-dimensional versions
\cite{Guan2013,Pagano2014,Decamp2016,Dobrzyniecki2020}.

\section*{Acknowledgments}
JHH acknowledges the financial support of the Graduate School for Integrative
Science \& Engineering at the National University of Singapore.
The Centre for Quantum Technologies is a Research Centre of Excellence funded
by the Ministry of Education and the National Research Foundation of
Singapore.  

\enlargethispage{1.0\baselineskip}

\end{document}